# Structural Modeling of DNA Loops in Lactose-Repressor

**S. Goyal[1] and N. C. Perkins**
University of Michigan, Mechanical Engineering
2350 Hayward, Ann Arbor, Michigan-48109-2125
Email. sgoyal@umich.edu

Experimental studies of DNA (Deoxy-ribo-Nucleic Acid) have established that its long-length scale structural mechanics play a crucial role in its biological functions including gene expression. For instance, long-length scale 'supercoiling' regulates the unwinding of the smaller-length scale helical coils of its constituent strands (the two sugar-phosphate chains) that enclose the genetic information (the base pairs), refer to Calladine et al. [1]. Many proteins also interact with DNA to deform it into loops (and supercoils) as crucial steps in promoting or repressing gene expression.

The goal of our research is to model DNA as a nonlinear elastic rod under two-axis bending and torsion to explore its long-length scale looping behavior. This goal is conceptually depicted below in Fig. 1. Loop formation in elastic rods is often initiated by instabilities under compression and/or torsion and accompanies nonlinear transitions to more energetically favorable equilibria. These large nonlinear deformations are governed by various structural properties of the rod including its bending and torsional stiffnesses, tension-torsion coupling (due to helical 'chirality'), and stress-free curvature. Furthermore, these properties are known to vary along the length of DNA because they depend on the base-pair sequence. The dynamics of a DNA strand is also influenced by its physical interactions with the surrounding fluid medium. For example, the fluid exerts hydrodynamic and thermal forces on DNA, and its ionic composition screens electrostatic repulsion of the negatively charged phosphate backbone of DNA.

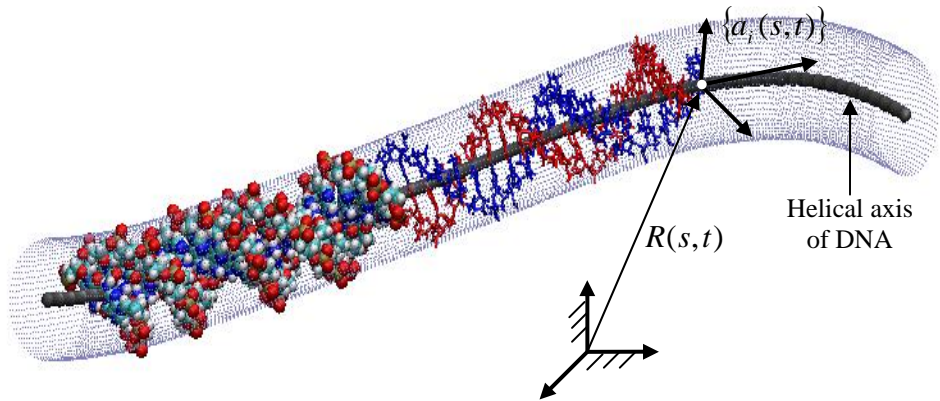

**Figure 1**: Structural Modeling DNA as a nonlinear elastic rod.

Our nonlinear dynamic rod model (described in Goyal et al. [2]) yields a set of nonlinear partial differential equations in time and space (along the rod's contour length). These equations need to be solved numerically under known initial and boundary conditions. We employ this computational rod model to simulate various examples of DNA supercoiling/looping, one of which is presented in Fig. 2.

---

[1] Corresponding author

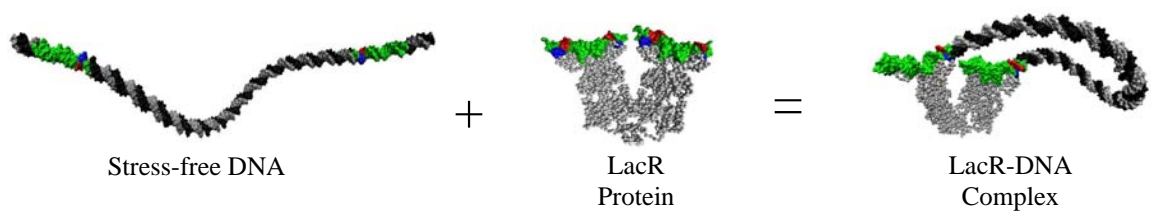

**Figure 2:** DNA loops in **Lac**tose-**R**epressor (LacR).

The activity of a lactose (sugar) producing gene in the bacterium *E.coli* is structurally controlled by a "**Lac**tose-**R**epressor" protein (abbreviated as LacR); refer to Fig. 2. The repressor protein binds to two sites ('operator' regions shown in green in Fig. 2) on this gene (DNA) to mechanically deform it into a loop. Protein-mediated looping of DNA is a crucial step (a 'biological switch') in many gene regulatory mechanisms and LacR is the most widely studied example of this phenomenon. The crystal structure of this DNA protein complex (described in Lewis et al. [3]) unfortunately cannot reveal the loop geometry, but it does provide clues about the boundary conditions for the loop. We use our computational rod model of DNA with these boundary conditions to predict the possible loop geometries and their deformation energies. The energetic cost of looping greatly influences the rate (or probability) of this mechano-chemical reaction (see again Fig. 2) and hence is an important parameter in the gene regulation. Both the geometry and energy of the loops strongly depend on the sequence-dependent material properties of DNA. Many ongoing experiments on the Lac-repressor protein are exploring the sequence-dependent behavior of looping by working with both the 'wild-type' (naturally occurring) gene as well as various man-made sequences that are designed to amplify individual structural properties. For instance, refer to the designed sequences of Mehta and Kahn [4] that introduce large intrinsic curvature in stress-free state. To complement these experiments, we explore through simulation the sensitivity of the loop geometry and energy to stress-free curvature, and chirality.